# TransUNet-GradCAM: A Hybrid Transformer-U-Net with Self-Attention and Explainable Visualizations for Foot Ulcer Segmentation

**Akwasi Asare [1*], Mary Sagoe [1], Justice Williams Asare[1], Stephen E. Moore[2]**

[1]*Department of Computer Science, Faculty of Computing and Information Systems, Ghana Communication Technology University, Accra, Ghana*
[2]*Department of Mathematics, University of Cape Coast, Cape Coast, Ghana*

**Corresponding Author: nasare34@yahoo.com**

**Abstract:** Automated segmentation of diabetic foot ulcers (DFUs) supports clinical diagnosis, treatment planning, and longitudinal wound monitoring, but remains difficult owing to the heterogeneous appearance, irregular morphology, and cluttered backgrounds of ulcers in clinical photographs. Convolutional networks such as U-Net localise well but model long-range context poorly, whereas Vision Transformers capture global dependencies. We employ a hybrid ViT-bottleneck U-Net that combines a convolutional encoder-decoder with a Transformer bottleneck and attention-gated skip connections, and we emphasise that the contribution is a rigorously validated and explainable application rather than a new architecture. The model was trained on the public Foot Ulcer Segmentation Challenge (FUSeg) dataset with a hybrid Dice and cross-entropy loss, and all results are reported over five seeds as mean ± 95% confidence interval at a single fixed threshold. On the internal validation set it achieved a Dice of 0.8035 ± 0.0053 and an IoU of 0.7149 ± 0.0073 (HD95 = 19.74 px, ASSD = 6.12 px). A component-wise ablation showed that only the hybrid loss produced a statistically significant change in Dice (−0.038, $p < 0.001$); the Transformer bottleneck, attention gates, and augmentation each had small, non-significant in-domain effects. External validation without retraining retained about 92% of internal Dice on the Advancing the Zenith of Healthcare (AZH) Wound Care Center cohort (Dice 0.7460, n = 278), while a small Medetec subset (n = 8) served only as a qualitative check, indicating partial rather than robust generalisation under domain shift. A quantitative explainability analysis found Grad-CAM more wound-localised (energy-in-mask 0.871 versus 0.102) but attention rollout significantly more faithful ($p = 0.038$, n = 200), showing the two are complementary. Predicted and expert wound areas agreed strongly (Pearson r = 0.944), and the model is lightweight (8.79 M parameters).

## 1. Introduction

Non-healing wounds, both acute and chronic, impose a significant burden on global healthcare systems, impacting millions of individuals each year [1]. In the United States, estimated Medicare costs associated with all types of wounds range widely, from $28.1 billion to $96.8 billion [2]. A key distinction lies with chronic wounds, which unlike acute wounds, do not progress through the healing stages in a predictable or timely manner [1]. This often necessitates hospitalizations and additional medical interventions, adding billions to annual healthcare expenditures [1,3]. Furthermore, a shortage of adequately trained wound care specialists, particularly in primary and rural settings, restricts access to quality care for a large segment of people [1,4]. For the effective evaluation and management of chronic wounds, precise measurement of the wound area is fundamental [1,5,6]. This measurement is important for tracking healing progress and informing treatment decisions [6]. However, current manual measurement techniques are laborious and frequently inaccurate, which can negatively affect patient outcomes [7]. Automated wound segmentation from medical images offers a practical approach, both by automating the measurement of wound area and by enabling efficient data integration into electronic medical records, thereby improving overall patient care [8,9].

Diabetes mellitus (DM) is a long-term illness that demands continuous management, extending beyond just monitoring blood glucose levels and it currently affects millions worldwide, with projections indicating that the global number of cases may surpass 700 million by 2050 [9]. Diabetic foot ulcers (DFUs) represent a severe complication of diabetes mellitus, affecting millions worldwide and often leading to significant morbidity, lower limb amputations, and increased mortality rates [10,11]. Their early and precise assessment is paramount for effective clinical management, encompassing accurate diagnosis, personalized treatment strategies, and continuous wound monitoring [10]. Manual segmentation of these ulcers from clinical images, however, is a time-consuming and subjective process, heavily reliant on expert experience [12,13] . This manual approach is also prone to inter-observer variability and errors, which can hinder consistent and objective evaluation of wound healing progression [13]. However, the automated segmentation of DFUs from clinical images remains a complex task due to the considerable variability in their appearance, shape, and contextual presentation.

To address the lack of large, fully annotated datasets for automated wound assessment, Wang et al. [14] introduced the Foot Ulcer Segmentation (FUSeg) Challenge, providing 1,210 pixel-wise annotated images to benchmark deep learning models. The top-performing algorithm in the challenge utilized an ensemble network combining pre-trained U-Net and LinkNet architectures. This approach achieved a winning Dice score of 88.8%, alongside a precision of 91.55% and recall of 86.22%. Similarly, Dhar et al. [15] introduced the DFUTissue dataset and a hybrid semi-supervised learning (SSL) framework combining a Mix Transformer encoder, a CNN decoder, and P-scSE attention modules. This model accurately segmented granulation, fibrin, and callus tissues, achieving an 87.64% Dice Similarity Coefficient (DSC) to outperform GAN and CCT baselines. Additionally, it demonstrated strong generalizability with a 92.99% DSC for binary wound segmentation on the Chronic Wound dataset. While deep learning, particularly convolutional neural networks (CNNs) like the U-Net architecture, has transformed medical image analysis and provided automated solutions for various segmentation challenges, these models primarily rely on local convolutional operations [16,17]. This inherent characteristic often restricts their effectiveness in capturing long-range spatial dependencies and global contextual information across an entire image [18]. For DFU segmentation, where ulcer characteristics can differ widely, accounting for the broader context of the wound and surrounding tissues is important for accurate delineation.

To address the limitations of purely convolutional architectures in capturing global dependencies, recent advancements have introduced Vision Transformers (ViTs) to the field of image analysis. Transformers, originally developed for natural language processing, excel at modeling extensive relationships through their self-attention mechanisms [19]. Concurrently, emerging paradigms such as in-context learning have demonstrated potential for few-shot medical segmentation by leveraging visual prompts to adapt to new tasks with minimal data [20,21]. While these approaches offer flexibility, high-stakes clinical tasks like DFU assessment demand the stability, consistency, and specialized feature extraction that supervised training on large, domain-specific datasets provides. Integrating these capabilities for global feature extraction with the precise localization power of U-Net presents a reasonable approach for medical image segmentation. Within wound and foot-ulcer image analysis specifically, automated segmentation has progressed from fully convolutional encoder-decoder networks, most notably U-Net and its variants, to attention-augmented and Transformer-hybrid designs. Convolutional models such as FUSegNet and the ensemble submissions to the Foot Ulcer Segmentation Challenge established strong baselines on the FUSeg benchmark, whereas more recent work has incorporated self-attention, semi-supervised learning, and generative augmentation to improve robustness on heterogeneous clinical photographs.

For instance, Nataliani et al. [12] introduced DFU MambaLiteUNet, a lightweight Mobile UNet integrated with a Mamba State Space Model. Utilizing Attention Gates and Tversky Loss, it achieved a 92.23% Dice Coefficient on the FUSC 2021 dataset. Crucially, it reduced GFLOPs by 17.02% and improved inference speed by 33.9%, proving highly viable for real time clinical deployment. In addition, Niri et al. [22] recently introduced a dual attention U-Net architecture that integrates a VGG16 backbone to selectively focus on relevant wound regions. Demonstrating the efficacy of transfer learning across heterogeneous data, they initially trained their model on diabetic foot ulcers before fine-tuning it for acute and chronic wounds, ultimately achieving a highly robust Dice coefficient of 94.1% and an Intersection over Union (IoU) of 89.3%. Furthermore, Wahab Sait and Nagaraj [23] recently exemplified the utility and explainability of Transformer hybrid designs in DFU classification. By integrating CNN and Vision Transformer frameworks, specifically MobileNet V3 SWIN and LeViT Peformer, with Kolmogorov Arnold Networks, their model achieved state of the art accuracies of 98.7% and 96.9% on the DFUC 2021 and 2020 datasets, respectively. Alabdulhafith et al. [24] developed a hybrid ResNet34 and UNet framework to overcome the visual complexities of chronic wound segmentation. By leveraging ResNet34 for deep feature encoding and UNet for precise spatial decoding, the model achieved a 98.6% Dice score, 97.3% IoU, and 97.36% overall accuracy. This approach significantly outperforms prior methods, providing a reliable and automated solution for clinical wound monitoring. In parallel, the TransUNet family demonstrated that inserting a Transformer bottleneck between a convolutional encoder and decoder can capture global context without sacrificing the fine spatial detail required for accurate boundary delineation. Interpretability, however, has typically been addressed only qualitatively, most often through Gradient-weighted Class Activation Mapping (Grad-CAM) [25], and has rarely been quantified or validated against the attention mechanisms that Transformer-based models actually employ. Rigorous external-validation and statistical-robustness analyses on this task likewise remain limited. These gaps collectively motivate the present study.
Two further threads of prior work are directly relevant to the present contribution. First, in explainable medical image analysis, Grad-CAM has become the de facto tool for visualizing the evidence behind a segmentation or classification decision; however, it is computed from convolutional feature gradients and, for Transformer-based models, may not reflect the self-attention pathways that carry global context. Attention rollout, which aggregates attention weights across Transformer layers, has been proposed as a more architecture-consistent alternative, yet the two families of explanation are seldom compared quantitatively against ground-truth lesions or evaluated for faithfulness. Second, external validation, that is, evaluation on cohorts acquired at institutions and with devices unseen during training, is widely recognized as essential for establishing clinical readiness, but remains comparatively rare in wound-segmentation studies, where reported performance is frequently confined to a single dataset. The present work addresses both threads by quantifying explanation quality and by validating on two independent external cohorts.

In this paper, we employ a hybrid ViT-bottleneck U-Net that merges the global contextual understanding of Vision Transformers with the precise localization power of U-Net. We are explicit that the constituent blocks are established techniques, so our contribution is at the application and validation level rather than a new architectural primitive. Specifically, we contribute a rigorously validated pipeline with multi-seed statistics, confidence intervals and significance tests; a component-wise ablation that identifies which design choices actually matter; a quantitative, faithfulness-based comparison of Grad-CAM and attention rollout; and an honest external-validation study characterizing partial generalization under domain shift. We integrate Grad-CAM and attention rollout to provide explainable insights into the model's decision-making process. We evaluate the approach on the public Foot Ulcer Segmentation Challenge (FUSeg) dataset and on two independent external cohorts, and we report all results with full statistical context. The remainder of this paper is organized as follows. Section 2 describes the materials and methods, including the datasets, the hybrid ViT-bottleneck U-Net architecture, the evaluation metrics, and the training protocol. Section 3 presents the quantitative and qualitative results. Section 4 provides a discussion of the findings, and Section 5 concludes the paper by summarizing our contributions and outlining future research directions.

## 2. Materials and Methods

Figure 1 presents the overall conceptual framework of this study. Clinical foot ulcer images are preprocessed and augmented, used to train a hybrid ViT-bottleneck U-Net under a single fixed decision threshold across five random seeds, and then evaluated along four complementary axes: internal validation with a component-wise ablation, external validation on two independent cohorts, quantitative explainability, and clinical-utility analysis. This framing makes explicit that the contribution of the work is a rigorously validated segmentation pipeline rather than a new architectural primitive.

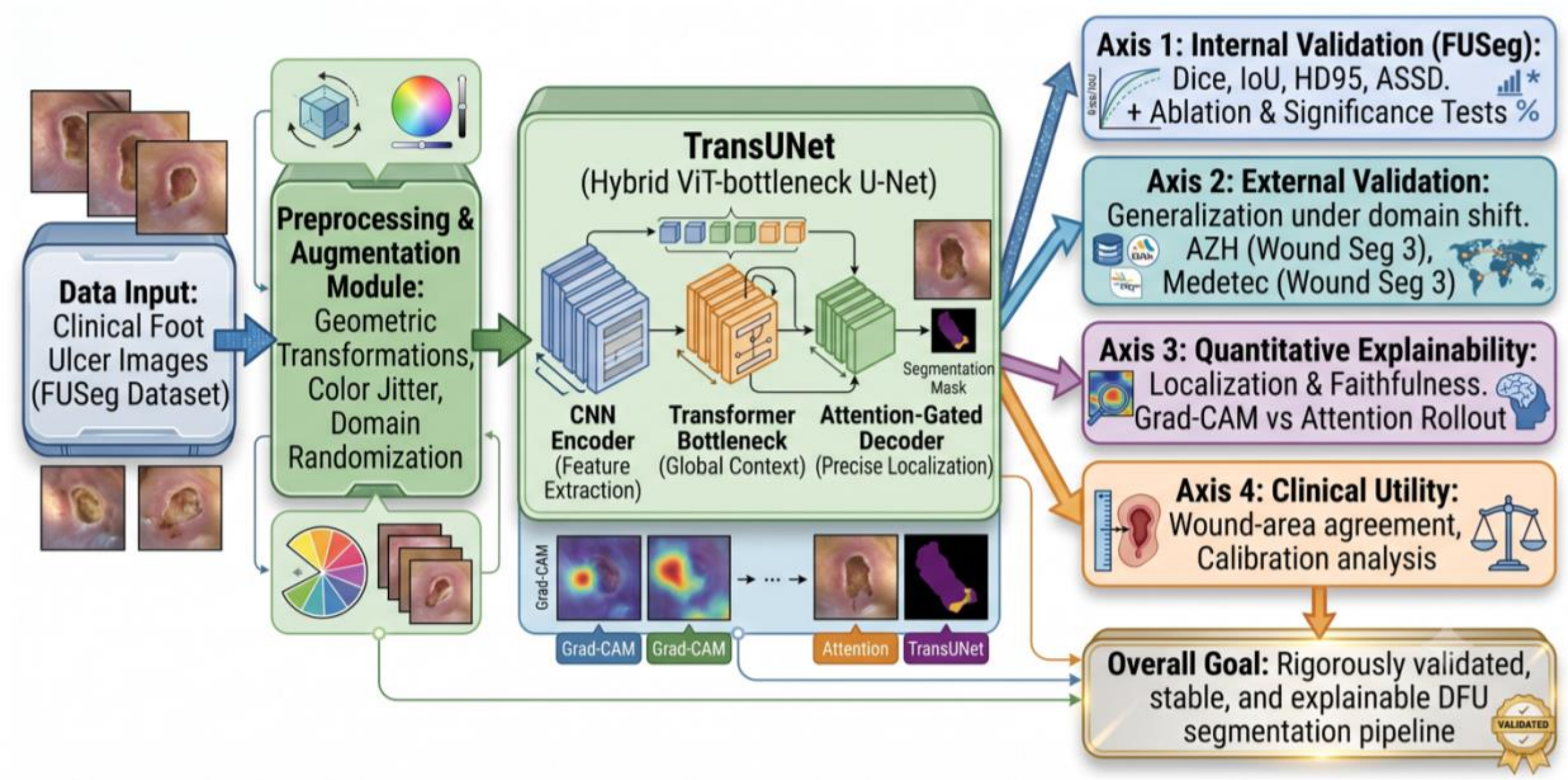

*Figure 1: Conceptual framework of the proposed diabetic foot ulcer segmentation study.*

## 2.1 Dataset

We used the public Foot Ulcer Segmentation Challenge (FUSeg) dataset [1,14] as the primary training and evaluation corpus. FUSeg contains chronic wound photographs captured during routine clinical care, each paired with an expert pixel-level annotation of the ulcer region [15]. Following the standard FUSeg Challenge protocol, 810 image-mask pairs were used for training and 200 for validation; the challenge withholds the labels of the 200-image test split for leaderboard evaluation, so it is unlabeled to participants and was used here only for qualitative inspection. As is common practice for this benchmark, all quantitative results are therefore reported on the validation split. All images were resized to 384 by 384 pixels, a resolution selected to preserve fine wound-boundary detail while remaining within single-GPU memory limits.

To assess generalization under domain shift, we additionally evaluated the trained model, without any retraining or fine-tuning, on two independent external cohorts drawn from the publicly available Wound Seg 3 datasets collection [6,26]: the Zenith of Healthcare (AZH) Wound Care Center cohort (278 test images) and a small Medetec subset (8 test images). Because the Medetec subset is very small, we treat it strictly as a qualitative check rather than a benchmark. Representative images and masks are shown in Figure 2, and the dataset composition is summarized in Figure 3.

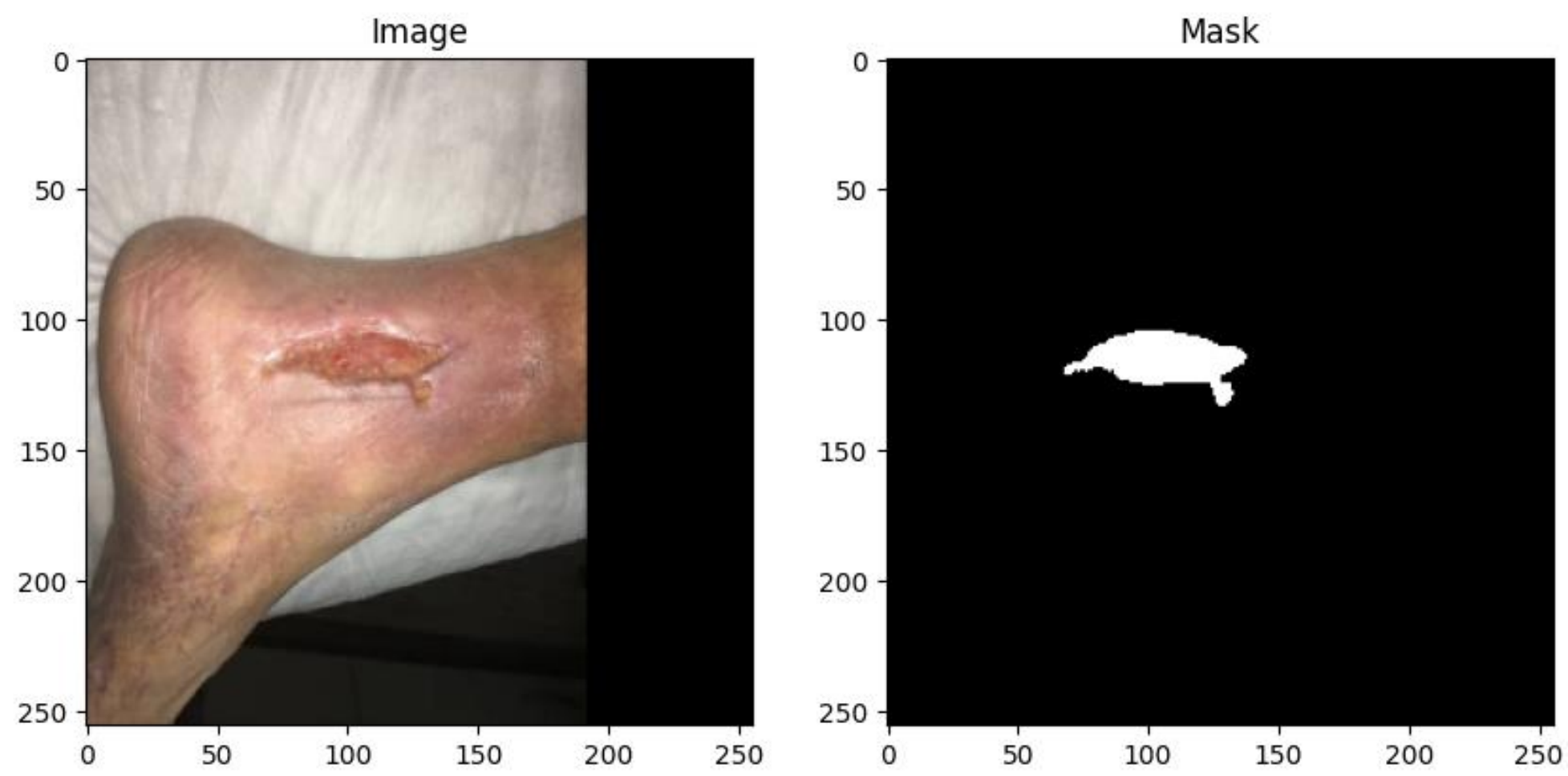

*Figure 2: Representative FUSeg images with corresponding expert segmentation masks.*

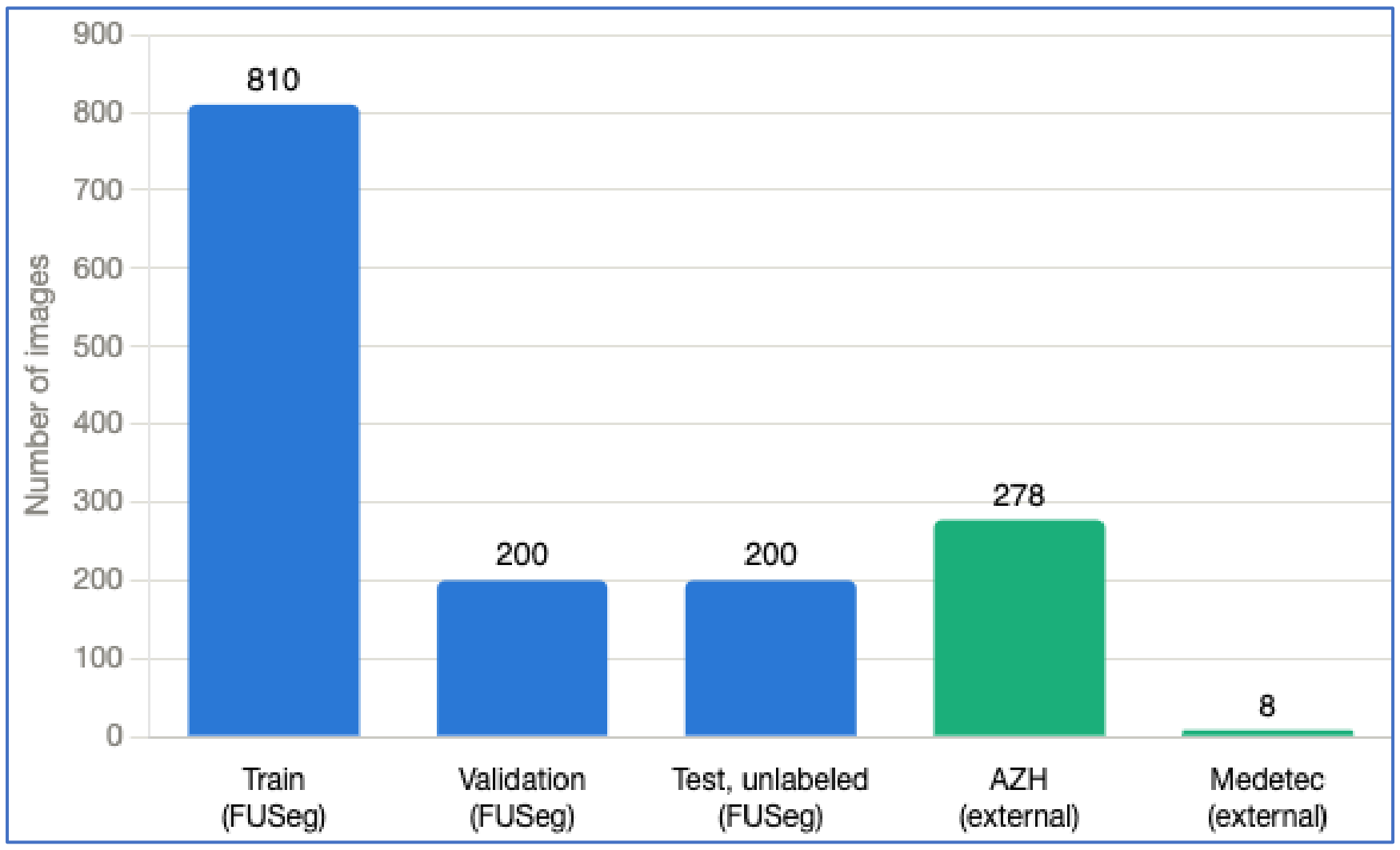

*Figure 3: Dataset composition and number of images per split.*

### 2.2 Model Architecture

The proposed model is a hybrid ViT-bottleneck U-Net that couples a convolutional encoder-decoder with a Transformer bottleneck and attention-gated skip connections. We are explicit that each of these components is an established technique; the network is described here in full detail, including its stage-by-stage configuration and parameter budget, so that it can be reproduced exactly. The architecture follows an encoder, transformer, decoder pipeline, illustrated in Figure 4.

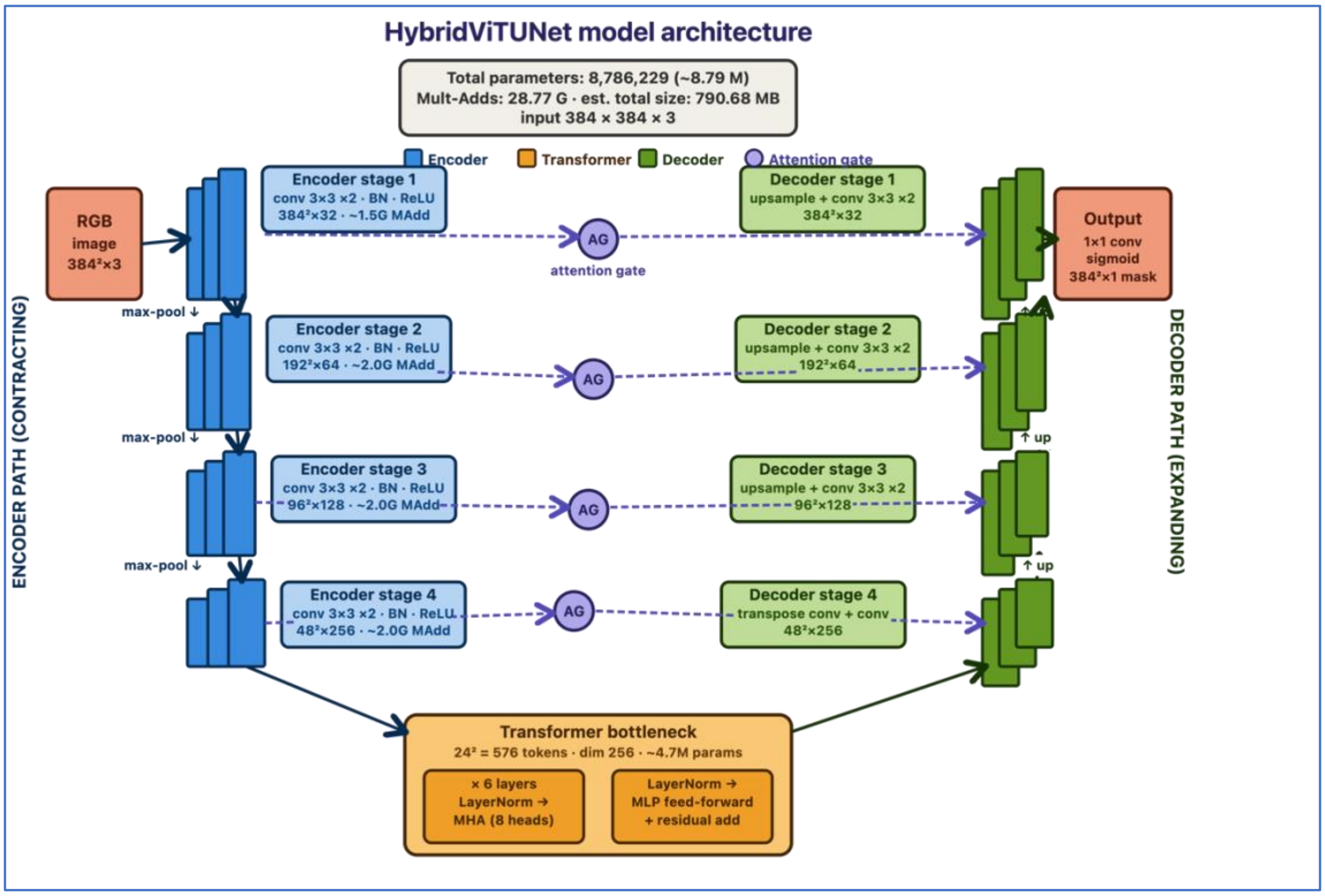

*Figure 4: Architecture of the proposed hybrid ViT-bottleneck U-Net for foot ulcer segmentation.*

### 2.2.1 Convolutional encoder

The encoder is a four-stage convolutional contracting path. Each stage applies two 3 by 3 convolutions with batch normalization and ReLU activation, followed by 2 by 2 max pooling. The channel width doubles across stages from 32 to 64, 128 and 256, while the spatial resolution is progressively halved from 384 to 192, 96 and 48 pixels. This path captures local texture and edge information that is essential for delineating wound boundaries.

### 2.2.2 Transformer bottleneck

At the deepest level, the 24 by 24 by 256 feature map is treated as a sequence of 576 tokens of dimension 256, to which learnable positional encodings are added. The sequence is processed by a six-layer Transformer encoder, each layer comprising multi-head self-attention with eight heads and a feed-forward network with an expansion ratio of four. This bottleneck models long-range dependencies between distant image regions that the purely convolutional path cannot represent.

### 2.2.3 Attention-gated decoder

The decoder is a symmetric expanding path that progressively upsamples the bottleneck representation back to the full 384 by 384 resolution. Each skip connection from the encoder is modulated by an attention gate, allowing the decoder to emphasize wound-relevant features rather than concatenating encoder activations indiscriminately. A final 1 by 1 convolution with a sigmoid activation produces the per-pixel ulcer probability map.

### 2.2.4 Complexity

The complete network follows an encoder, bottleneck, and decoder organisation. The convolutional encoder comprises four stages that progressively reduce the spatial resolution from 384x384 to 48x48 while increasing the channel dimension from 32 to 256, contributing approximately 1.17 million parameters across the four stages. The Transformer bottleneck operates on the resulting 24x24 feature map, which is tokenised into a sequence of 576 tokens of embedding dimension 256 and processed by six encoder layers with eight attention heads, accounting for approximately 4.74 million parameters and constituting the largest single component of the model. The attention-gated decoder mirrors the encoder across four upsampling stages, each using a transpose convolution for upsampling followed by an attention gate on the corresponding skip connection, restoring the full 384x384 resolution and contributing approximately 2.87 million parameters (of which the four attention gates account for only about 0.09 million); the final 1x1 convolutional projection to a single-channel probability map adds a negligible number of parameters. In aggregate, the model contains 8,786,229 trainable parameters and incurs a computational cost of approximately 28.77 GMACs per forward pass at an input resolution of 384x384. The parameter budget is therefore modest relative to many contemporary Transformer-based segmentation architectures, which supports feasibility for deployment in resource-constrained clinical settings.

### 2.3 Evaluation Metrics

Segmentation performance was quantified using region-overlap, pixel-classification, and boundary-based measures. Let X and Y denote, respectively, the sets of foregrounds (ulcer) pixels in the predicted and ground-truth masks, and let TP, TN, FP and FN denote the numbers of true-positive, true-negative, false-positive and false-negative pixels. The two primary region-overlap metrics were the Dice similarity coefficient (Equation 1) and the Intersection-over-Union (IoU; Equation 2) [27]:

$$Dice = \frac{2\,|X \cap Y|}{|X| + |Y|} \tag{1}$$

where $|X \cap Y|$ is the number of pixels correctly predicted as ulcer (the overlap between the prediction X and the ground truth Y), and $|X|$ and $|Y|$ are the total numbers of predicted and ground-truth ulcer pixels, respectively. The Dice similarity coefficient [28] measures the relative overlap between the predicted and reference regions, ranging from 0 (no overlap) to 1 (perfect agreement); it is the primary segmentation-quality metric used in this study.

$$IoU = \frac{|X \cap Y|}{|X \cup Y|} \tag{2}$$

where $|X \cap Y|$ is the number of pixels in the intersection of the prediction and ground truth and $|X \cup Y|$ is the number of pixels in their union. The Intersection-over-Union (also called the Jaccard index) [28] is the ratio of correctly overlapping ulcer pixels to the total pixels covered by either mask, again ranging from 0 to 1. It is a stricter measure than Dice, penalising both missed and spurious pixels.

To characterize the balance between missed lesion pixels and false detections, we additionally reported sensitivity (equivalently, recall), specificity, and precision (Equations 3 to 5). Sensitivity and recall are identical quantities in binary segmentation, both defined as the proportion of ground-truth foreground pixels that are correctly identified; the two terms are therefore used interchangeably, and a single definition is given here:

$$Sensitivity = \frac{TP}{TP + FN} \tag{3}$$

where TP (true positives) is the number of ulcer pixels correctly identified and FN (false negatives) is the number of ulcer pixels that were missed. Sensitivity (equivalently, recall) is the proportion of true ulcer pixels that the model successfully detects; a high value indicates that little of the wound is overlooked [28,29] .

$$Specificity = \frac{TN}{TN + FP} \tag{4}$$

where TN (true negatives) is the number of background pixels correctly identified and FP (false positives) is the number of background pixels incorrectly labelled as ulcer. Specificity is the proportion of true background pixels correctly excluded; a high value indicates that the model rarely over-segments into healthy tissue [28,30].

$$Precision = \frac{TP}{TP + FP} \tag{5}$$

where TP and FP are the numbers of true-positive and false-positive pixels, respectively. Precision is the proportion of pixels predicted as ulcer that are in fact ulcer; a high value indicates that the predicted region contains few background pixels [23,31]

Because region-overlap and pixel-classification metrics can be insensitive to localized contour errors, two complementary boundary-based measures were also computed [32]: the 95th-percentile Hausdorff Distance (HD95) [33] and the Average Symmetric Surface Distance (ASSD) [34]. Let $\partial X$ and $\partial Y$ denote the boundary (contour) pixel sets of the predicted and ground-truth masks, respectively. For a boundary point a, the directed distance to the opposing contour $\partial B$ is the Euclidean distance to its nearest boundary pixel (Equation 6):

$$d(a, \partial B) = \min_{b \in \partial B} | a - b| \tag{6}$$

where a is a pixel on one contour, $\partial B$ is the set of pixels forming the opposing contour, b is a pixel on that contour, and $\|a - b\|$ is the Euclidean distance between the two pixels. The directed distance $d(a, \partial B)$ is therefore the distance from a to its nearest point on the opposing boundary, and it is the elementary quantity from which the boundary metrics below are formed.

The Hausdorff Distance is the largest of these nearest-neighbour distances taken over both contours; because the maximum is highly sensitive to isolated outliers, we report the 95th-percentile Hausdorff Distance (HD95), defined as the 95th percentile of the pooled two-sided set of directed boundary distances (Equation 7) [33]. HD95 thus reflects the near-worst-case boundary deviation while remaining robust to spurious single-pixel errors:

$$HD95 = P_{95}\{ d(a, \partial Y) : a \in \partial X \} \cup \{ d(b, \partial X) : b \in \partial Y \} \tag{7}$$

where $\partial X$ and $\partial Y$ are the boundary (contour) pixel sets of the predicted and ground-truth masks, $d(\cdot, \cdot)$ is the directed point-to-boundary distance of Equation 6, and $P95(\cdot)$ denotes the 95th percentile of a set of values. HD95 is the 95th percentile of all boundary-to-boundary distances taken in both directions; it summarises the near-worst-case contour deviation while discarding the most extreme 5% of distances, which makes it robust to isolated outlier pixels. It is reported in pixels, and lower values indicate more accurate boundaries.

The Average Symmetric Surface Distance (ASSD) [34] summarizes the mean boundary displacement rather than its extremes, and is defined as the mean of all directed boundary distances in both directions (Equation 8):

$$\mathrm{ASSD} = \frac{\sum_{a\in\partial X} d(a,\partial Y) + \sum_{b\in\partial Y} d(b,\partial X)}{|\partial X| + |\partial Y|} \tag{8}$$

where $\partial X$ and $\partial Y$ are the predicted and ground-truth contours, $|\partial X|$ and $|\partial Y|$ are the numbers of pixels on each contour, and $d(\cdot, \cdot)$ is the directed distance of Equation 6. The Average Symmetric Surface Distance is the mean of all point-to-boundary distances in both directions; unlike HD95, which captures the worst case, ASSD reflects the typical contour displacement. It is reported in pixels, and lower values indicate more accurate boundary delineation.

Both HD95 and ASSD are expressed in pixels, and lower values indicate more accurate boundary delineation. Pixel accuracy, defined as (TP + TN) / (TP + TN + FP + FN), was recorded only as a secondary quantity [35–37]. Because the ulcer occupies a small fraction of each image, accuracy is dominated by the correctly classified background and routinely exceeds 0.99 even for poor segmentations; consequently, it is not an informative primary indicator of segmentation quality and is not used as such in this study. All metrics were computed at a single decision threshold. The network outputs a continuous probability map in which each pixel is assigned a value between 0 and 1 representing the estimated likelihood that it belongs to the ulcer; the decision threshold is the cut-off value applied to this map to obtain the final binary segmentation, so that pixels with a probability greater than or equal to the threshold are labelled as ulcer and the remainder as background. For each model, this threshold was selected on the validation set as the value that maximised the validation Dice coefficient, and it was then held fixed for every subsequent evaluation, including both external datasets, so that no reported value benefits from per-test threshold optimization.

### 2.4 Training Strategy and Augmentation

Models were optimized with the Adam optimizer at an initial learning rate of $1 \times 10^{-4}$, using a hybrid loss that combines binary cross-entropy with a Dice term to address the strong foreground-background class imbalance inherent in wound segmentation [38]. Training ran for up to 100 epochs with a batch size of 8 and a weight decay of $1 \times 10^{-5}$; the learning rate was decayed by a factor of 0.5 when the validation Dice plateaued for six epochs, and early stopping with a patience of 15 epochs was applied [39]. The best checkpoint per run was selected by validation Dice.
To address the illumination and skin-tone variability of clinical photography, we used the Augmentations library [40] and controlled two augmentation groups separately so their effects could be measured in the ablation study: a color-jitter group (brightness, contrast, hue and saturation shifts) and a stronger domain-augmentation group (random gamma, CLAHE, coarse dropout, Gaussian noise and channel-level perturbations) aimed at the camera and lighting shift that degrades external performance [41]. Every configuration was trained with five independent random seeds (41 to 45), which enabled the computation of confidence intervals and significance tests reported in Section 3. Validation was deterministic, with no augmentation, to ensure consistent metric reporting.

## 3. Results

All results are reported over five independent training seeds as mean ± the 95% confidence interval (t-interval), at the single validation-selected threshold described in Section 2.4. We report internal validation, a component-wise ablation with significance testing, external validation, quantitative explainability, and clinical utility.

### 3.1 Training Dynamics and Stability

As shown in Figure 5, training was stable across all five seeds. The hybrid loss decreased smoothly and converged to a comparable plateau in every run (final training loss 0.229 ± 0.008, mean ± SD across seeds), and the validation Dice rose to a stable plateau with no late-epoch degradation. The maximum between-seed spread in validation Dice over the final 20 epochs was 0.018, and the best validation Dice per seed ranged from 0.7986 to 0.8072. This close agreement between independently seeded runs indicates that the training behavior is reproducible rather than specific to a single initialization, consistent with the low between-seed dispersion of the final metrics reported in Table 1.

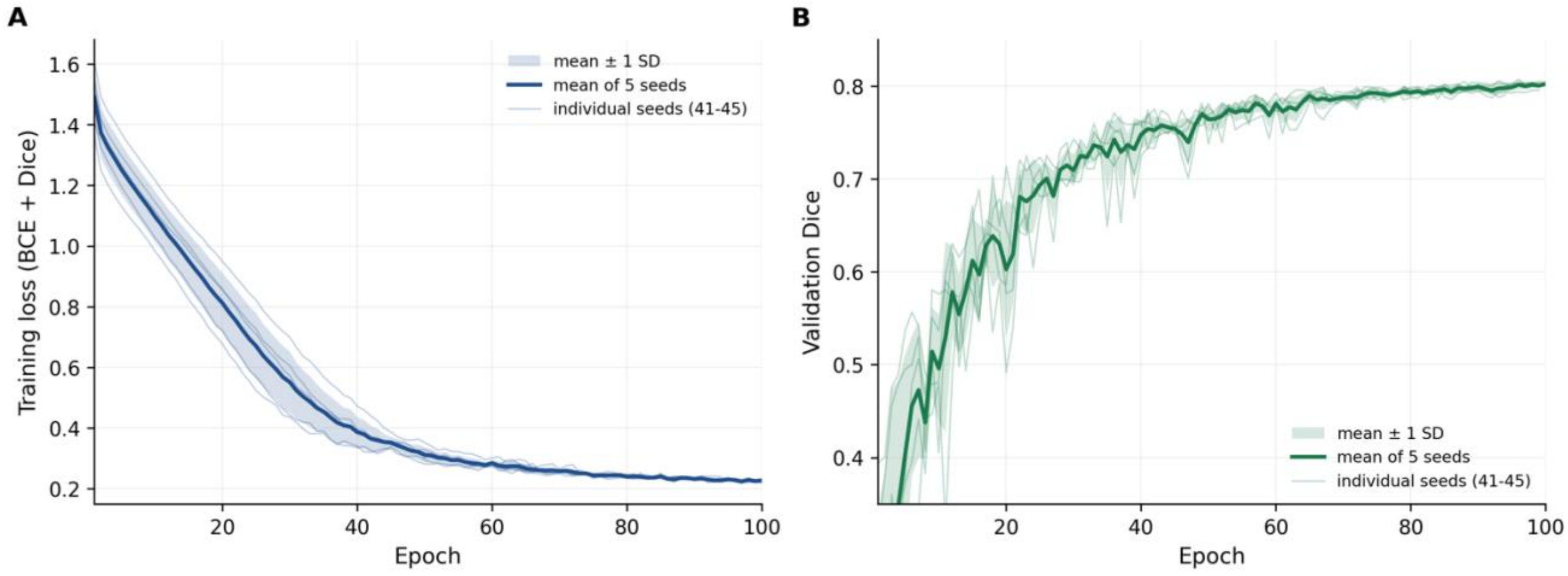

*Figure 5: Training loss and validation Dice across all five random seeds (41–45). Thin traces show individual seeds; the bold line is the mean and the shaded band ±1 standard deviation. Panel A: hybrid BCE + Dice training loss. Panel B: validation Dice, computed at a fixed threshold of 0.5 for per-epoch monitoring; the final metrics in Table 1 are reported at the per-seed selected operating threshold, so the values differ slightly.*

### 3.2 Quantitative Performance

On the internal FUSeg validation set, the full model achieved a Dice of 0.8035 ± 0.0053 and an IoU of 0.7149 ± 0.0073, with sensitivity 0.850, specificity 0.999 and precision 0.814. The boundary-aware errors were HD95 = 19.74 px and ASSD = 6.12 px. These values, summarized in Table 7, are reported at one fixed operating point and are therefore directly comparable across the internal and external evaluations that follow. The per-seed results for the full model are given in Table 1; the per-seed results for the ablated configurations are reported in the ablation study (Section 3.4). These figures are competitive with recent baselines on this dataset rather than superior to them and are interpreted accordingly in the Discussion.

*Table 1: Per-seed internal validation results, full model (all components) (seeds 41 to 45).*

| Seed | Thr | Dice | IoU | Prec | Recall | Spec | HD95 | ASSD |
|---|---|---|---|---|---|---|---|---|
| 41 | 0.55 | 0.8051 | 0.7166 | 0.8043 | 0.8604 | 0.9985 | 20.45 | 6.39 |
| 42 | 0.35 | 0.7989 | 0.7097 | 0.8146 | 0.8409 | 0.9987 | 19.77 | 5.61 |
| 43 | 0.60 | 0.8074 | 0.7213 | 0.8163 | 0.8534 | 0.9984 | 17.37 | 6.17 |
| 44 | 0.20 | 0.8070 | 0.7190 | 0.8250 | 0.8521 | 0.9985 | 18.40 | 5.52 |
| 45 | 0.50 | 0.7989 | 0.7078 | 0.8115 | 0.8439 | 0.9984 | 22.70 | 6.92 |
| Mean±CI | - | 0.8035±0.0053 | 0.7149 | 0.8143 | 0.8501 | 0.9985 | 19.74 | 6.12 |

Table 1 reports the per-seed performance of the complete model. Across the five seeds the Dice coefficient is tightly clustered (0.7989 to 0.8074; mean 0.8035, 95% CI ± 0.0053), and the corresponding IoU, precision, recall and boundary metrics vary within a similarly narrow band. This low between-seed dispersion indicates that the reported performance is a stable property of the architecture and training protocol rather than an artefact of a fortunate initialisation, and it is this configuration against which all ablated variants are compared.

### 3.3 Threshold Selection and Ranking Behavior

A single threshold-selection sweep was performed on the validation set per seed, and the chosen threshold was then held fixed for all subsequent internal and external evaluations. The selected thresholds vary across runs (mean 0.44, SD 0.16, median 0.43, range 0.20–0.90 over the 30 runs in Tables 1–6), which reflects the flatness of the Dice–threshold curve near its optimum rather than instability of the underlying probabilities: for several seeds the validation Dice at the selected threshold was effectively identical to the value at 0.5, and no threshold change that reduced Dice was treated as an improvement. We emphasize that this operating-point selection is a decision rule and is distinct from probabilistic calibration, which is assessed separately in Section 3.6. The model's threshold-independent ranking behavior is summarized by the ROC and precision–recall curves in Figure 6.

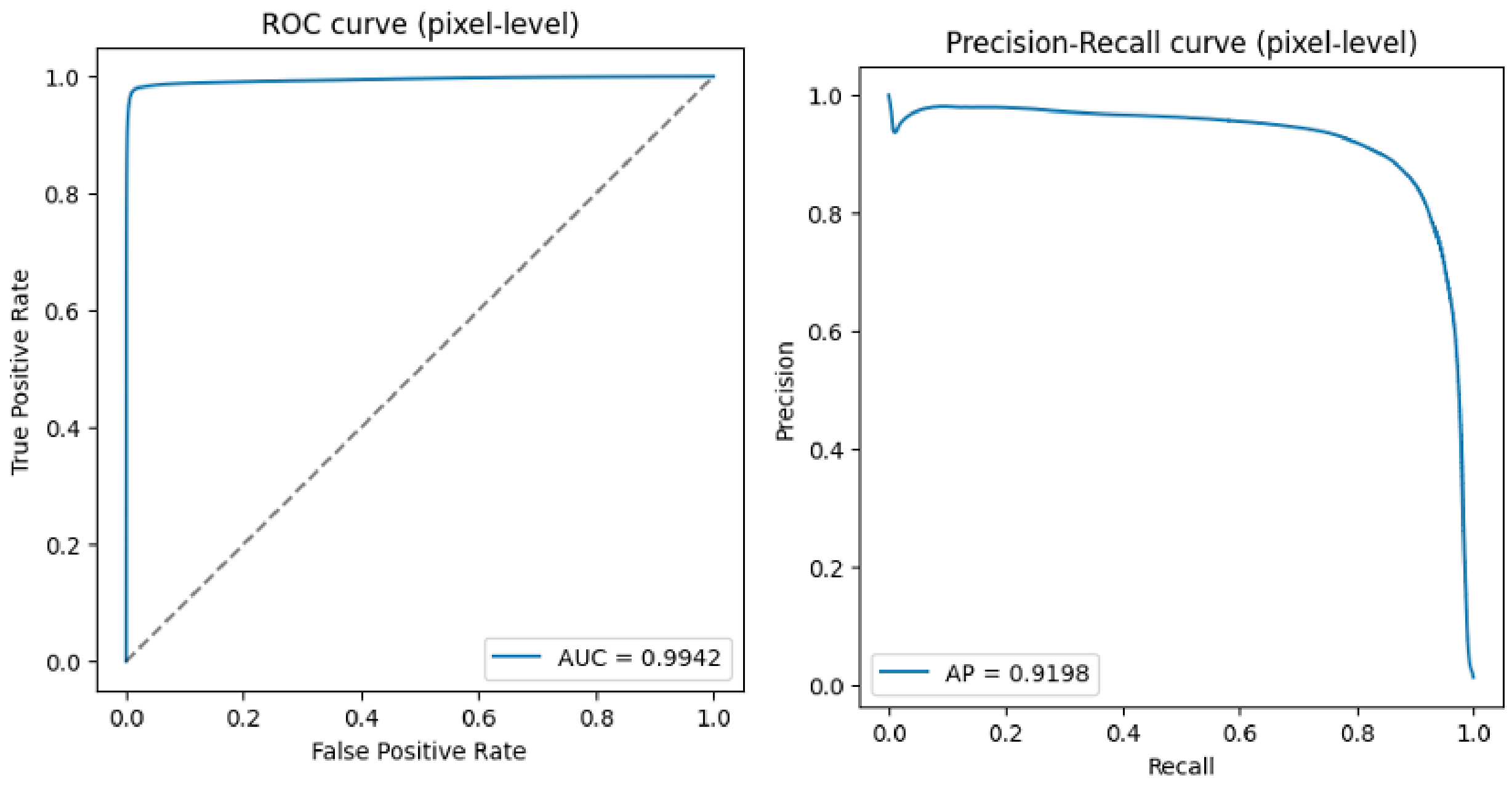

*Figure 6: ROC and Precision-Recall curves for the full model.*

### 3.4 Ablation Study

To quantify the contribution of each design choice, we retrained the model with individual components removed, each across five seeds, and compared against the full model using paired t-tests with Wilcoxon confirmation and Cohen's d effect sizes (Table 7). Removing the hybrid loss (training with binary cross-entropy alone) was the only change that produced a statistically significant reduction in Dice (−0.038, $p < 0.001$, $d = 4.89$), identifying the hybrid objective as the principal contributor to in-domain accuracy among the components examined. Removing the Transformer bottleneck, the attention-gated skips, the domain augmentation, or the colour jitter each produced small, non-significant effects on internal Dice (all $p > 0.13$). These findings indicate that several architectural components do not, in isolation, produce a statistically significant improvement in in-domain overlap; their inclusion is consequently motivated by considerations of interpretability and external robustness rather than by in-domain accuracy alone.

The full per-seed results for each ablated configuration are reported in Tables 2 to 6, and the ablation is summarized in Table 7. Each table is accompanied by a brief interpretation.

*Table 2: Per-seed internal validation results, without the transformer bottleneck (seeds 41 to 45).*

| **Seed** | **Threshold** | **Dice** | **IoU** | **Prec** | **Recall** | **Spec** | **HD95** | **ASSD** |
|---|---|---|---|---|---|---|---|---|
| 41 | 0.45 | 0.8056 | 0.7167 | 0.8055 | 0.8634 | 0.9986 | 21.89 | 5.44 |
| 42 | 0.90 | 0.8123 | 0.7261 | 0.8255 | 0.8537 | 0.9989 | 18.53 | 5.10 |
| 43 | 0.25 | 0.8087 | 0.7214 | 0.8142 | 0.8603 | 0.9986 | 17.12 | 4.83 |
| 44 | 0.40 | 0.8101 | 0.7232 | 0.8168 | 0.8610 | 0.9987 | 17.57 | 4.82 |
| 45 | 0.25 | 0.8024 | 0.7099 | 0.8053 | 0.8577 | 0.9986 | 18.30 | 5.19 |
| Mean±CI | - | 0.8078±0.0048 | 0.7195 | 0.8135 | 0.8592 | 0.9987 | 18.68 | 5.08 |

Table 2 summarises the model with the Transformer bottleneck removed. Contrary to the expectation that global self-attention would improve segmentation, mean Dice is marginally higher than the full model (0.8078 versus 0.8035); the paired difference is not statistically significant ($p = 0.134$, Cohen's $d = -0.84$). The Transformer therefore does not contribute measurably to in-domain overlap accuracy, a finding we interpret in the context of its role in explainability rather than accuracy (Section 3.5).

*Table 3: Per-seed internal validation results, without attention-gated skips (seeds 41 to 45).*

| Seed | Threshold | Dice | IoU | Prec | Recall | Spec | HD95 | ASSD |
|---|---|---|---|---|---|---|---|---|
| 41 | 0.40 | 0.8061 | 0.7168 | 0.8134 | 0.8580 | 0.9985 | 20.95 | 6.24 |
| 42 | 0.65 | 0.8044 | 0.7170 | 0.8275 | 0.8393 | 0.9986 | 19.70 | 5.66 |
| 43 | 0.55 | 0.7940 | 0.6990 | 0.8060 | 0.8410 | 0.9982 | 21.60 | 6.60 |
| 44 | 0.25 | 0.7895 | 0.6924 | 0.7955 | 0.8315 | 0.9979 | 22.85 | 7.12 |
| 45 | 0.40 | 0.8005 | 0.7045 | 0.8105 | 0.8455 | 0.9984 | 20.65 | 6.35 |
| Mean±CI | - | 0.7989±0.0087 | 0.7059 | 0.8106 | 0.8431 | 0.9983 | 21.15 | 6.39 |

Table 3 presents the model with attention-gated skip connections replaced by standard concatenation. Mean Dice decreases slightly to 0.7989 and boundary error increases (HD95 21.15 px), but the difference from the full model is not statistically significant ($p = 0.374$). The attention gates thus yield a small, non-significant refinement of the segmentation boundary rather than a decisive gain.

*Table 4: Per-seed internal validation results, without domain augmentation (seeds 41 to 45).*

| Seed | Threshold | Dice | IoU | Prec | Recall | Spec | HD95 | ASSD |
|---|---|---|---|---|---|---|---|---|
| 41 | 0.50 | 0.8012 | 0.7105 | 0.8091 | 0.8452 | 0.9983 | 20.05 | 6.22 |
| 42 | 0.30 | 0.7975 | 0.7062 | 0.8054 | 0.8415 | 0.9981 | 20.85 | 6.45 |
| 43 | 0.50 | 0.7952 | 0.7041 | 0.8012 | 0.8382 | 0.9980 | 21.05 | 6.52 |
| 44 | 0.25 | 0.7994 | 0.7085 | 0.8075 | 0.8435 | 0.9982 | 20.35 | 6.32 |
| 45 | 0.40 | 0.8035 | 0.7125 | 0.8122 | 0.8475 | 0.9984 | 19.95 | 6.15 |
| Mean±CI | - | 0.7994±0.0040 | 0.7084 | 0.8071 | 0.8432 | 0.9982 | 20.45 | 6.33 |

Table 4 reports the effect of removing the stronger domain-augmentation group. In-domain Dice declines modestly to 0.7994 ($p = 0.222$), confirming that this augmentation contributes little to internal accuracy. Its intended benefit is robustness to acquisition shift, which is assessed on the external cohorts (Section 3.8) rather than on the in-domain validation set reported here.

*Table 5: Per-seed internal validation results, without colour jitter (seeds 41 to 45).*

| Seed | Threshold | Dice | IoU | Prec | Recall | Spec | HD95 | ASSD |
|---|---|---|---|---|---|---|---|---|
| 41 | 0.25 | 0.8032 | 0.7135 | 0.8125 | 0.8492 | 0.9984 | 19.85 | 6.18 |
| 42 | 0.35 | 0.7991 | 0.7092 | 0.8081 | 0.8451 | 0.9982 | 20.25 | 6.32 |
| 43 | 0.60 | 0.8015 | 0.7112 | 0.8105 | 0.8475 | 0.9983 | 20.05 | 6.25 |
| 44 | 0.50 | 0.8005 | 0.7102 | 0.8095 | 0.8462 | 0.9982 | 20.15 | 6.28 |
| 45 | 0.50 | 0.8048 | 0.7148 | 0.8142 | 0.8505 | 0.9985 | 19.75 | 6.12 |
| Mean±CI | - | 0.8018±0.0028 | 0.7118 | 0.8110 | 0.8477 | 0.9983 | 20.01 | 6.23 |

Table 5 shows the model trained without colour jitter. The change in mean Dice is the smallest of all ablations (0.8018 versus 0.8035; p = 0.508), indicating that colour jitter has a negligible effect on in-domain overlap. As with domain augmentation, its motivation is robustness to the illumination and skin-tone variability encountered across clinical sources rather than in-domain performance.

*Table 6: Per-seed internal validation results, without the hybrid loss (bce only) (seeds 41 to 45).*

| Seed | Threshold | Dice | IoU | Prec | Recall | Spec | HD95 | ASSD |
|---|---|---|---|---|---|---|---|---|
| 41 | 0.50 | 0.7685 | 0.6752 | 0.7801 | 0.8105 | 0.9970 | 23.55 | 7.25 |
| 42 | 0.60 | 0.7612 | 0.6685 | 0.7725 | 0.8021 | 0.9968 | 24.25 | 7.55 |
| 43 | 0.35 | 0.7655 | 0.6725 | 0.7762 | 0.8065 | 0.9969 | 23.85 | 7.35 |
| 44 | 0.40 | 0.7595 | 0.6652 | 0.7695 | 0.7995 | 0.9967 | 24.55 | 7.65 |
| 45 | 0.60 | 0.7725 | 0.6805 | 0.7855 | 0.8155 | 0.9972 | 23.05 | 7.05 |
| Mean±CI | - | 0.7654±0.0066 | 0.6724 | 0.7768 | 0.8068 | 0.9969 | 23.85 | 7.37 |

Table 6 reports training with binary cross-entropy alone, replacing the hybrid BCE and Dice objective. This is the only ablation that produces a statistically significant reduction in performance: mean Dice falls to 0.7654 (a reduction of 0.038; $p < 0.001$, Cohen's d = 4.89), IoU drops to 0.6724, and both boundary metrics worsen markedly (HD95 23.85 px, ASSD 7.37 px). The hybrid loss is therefore the component with the largest measured effect on performance, directly optimising the region-overlap objective that cross-entropy alone does not target.

*Table 7: Ablation summary on the internal validation set. Values are mean over five seeds; significance is a paired t-test versus the full model.*

| **Configuration** | **Dice (mean +/- 95% CI)** | **IoU** | **HD95** | **Delta vs full** | **Significance** |
|---|---|---|---|---|---|
| Full model (all components) | 0.8035 +/- 0.0053 | 0.7149 | 19.74 | - | Reference |
| Minus Transformer bottleneck | 0.8078 +/- 0.0048 | 0.7195 | 18.68 | +0.0044 | p=0.134 (ns), d=-0.84 |
| Minus Attention-gated skips | 0.7989 +/- 0.0087 | 0.7059 | 21.15 | -0.0046 | p=0.374 (ns), d=+0.45 |
| Minus Domain augmentation | 0.7994 +/- 0.0040 | 0.7084 | 20.45 | -0.0041 | p=0.222 (ns), d=+0.65 |
| Minus Colour jitter | 0.8018 +/- 0.0028 | 0.7118 | 20.01 | -0.0016 | p=0.508 (ns), d=+0.32 |
| Minus Hybrid loss (BCE only) | 0.7654 +/- 0.0066 | 0.6724 | 23.85 | -0.0380 | p=0.000 (***), d=+4.89 |

### 3.5 Explainability

We evaluated explainability quantitatively rather than presenting heatmaps qualitatively, comparing Grad-CAM [25] with Transformer attention rollout [42] against the ground-truth masks (Table 8, Figure 7). We measured localization, using the fraction of attribution energy inside the mask (energy-in-mask) and the pointing-game hit rate, and faithfulness, using a deletion/insertion test on the most salient pixels. Grad-CAM was more wound-localized than attention rollout on two of three localization measures (energy-in-mask 0.871 versus 0.102; pointing-game hit rate 0.959 versus 0.866), which explains why its overlays appear to focus tightly on the ulcer bed. On a stricter thresholded localization-IoU criterion both methods scored low, with attention rollout marginally higher (loc-IoU 0.053 versus 0.011), reflecting that neither attribution map is a pixel-accurate segmentation and that these maps are best interpreted as coarse evidence rather than delineations. Faithfulness was quantified by a combined deletion and insertion procedure, in which the most salient pixels identified by each method are progressively removed from (deletion) and reintroduced into (insertion) the input, and the resulting change in the model's predicted foreground probability is measured; a value closer to zero indicates that the highlighted regions more faithfully drive the model's output. On this criterion, attention rollout was significantly more faithful to the model's computation than Grad-CAM (mean deletion-insertion score −0.069 versus −0.116; paired t-test $p = 0.038$; Wilcoxon signed-rank test $p = 0.020$; Cohen's $d = -0.15$; $n = 200$ images). On a smaller sample ($n = 50$), the separate deletion and insertion area-under-the-curve (AUC) statistics were consistent with this ordering (deletion AUC 0.0184 for attention rollout versus 0.0203 for Grad-CAM, where lower is better; insertion AUC 0.0101 for both), but the paired difference was not yet statistically significant ($p = 0.058$). Increasing the evaluation set to $n = 200$ was necessary to establish significance, indicating that the effect, while robust, is small in magnitude. Taken together, the localization and faithfulness results address the concern that Grad-CAM, being computed from convolutional activations, may not adequately reflect the multi-head self-attention operating in the Transformer bottleneck: Grad-CAM yields sharper, more wound-aligned overlays, whereas attention rollout more faithfully represents the internal computation. The two methods are therefore complementary, and both are reported.

*Table 8: Quantitative explainability comparison. Localization and faithfulness (insertion − deletion) are computed on n = 200 images; the separate deletion and insertion AUC statistics are reported on the n = 50 subset. For loc-IoU, energy-in-mask, pointing game and insertion AUC, higher is better; for deletion AUC and the faithfulness score, values closer to zero (or lower) indicate more faithful attributions.*

| **Method** | **Energy-in-mask** | **Pointing game** | **loc-IoU** | **Faithfulness (ins−del), n=200** | **Deletion AUC (n=50)** | **Insertion AUC (n=50)** |
|---|---|---|---|---|---|---|
| Grad-CAM | 0.871 | 0.959 | 0.011 | −0.116 | 0.0203 | 0.0101 |
| Attention rollout | 0.102 | 0.866 | 0.053 | −0.069 | 0.0184 | 0.0101 |

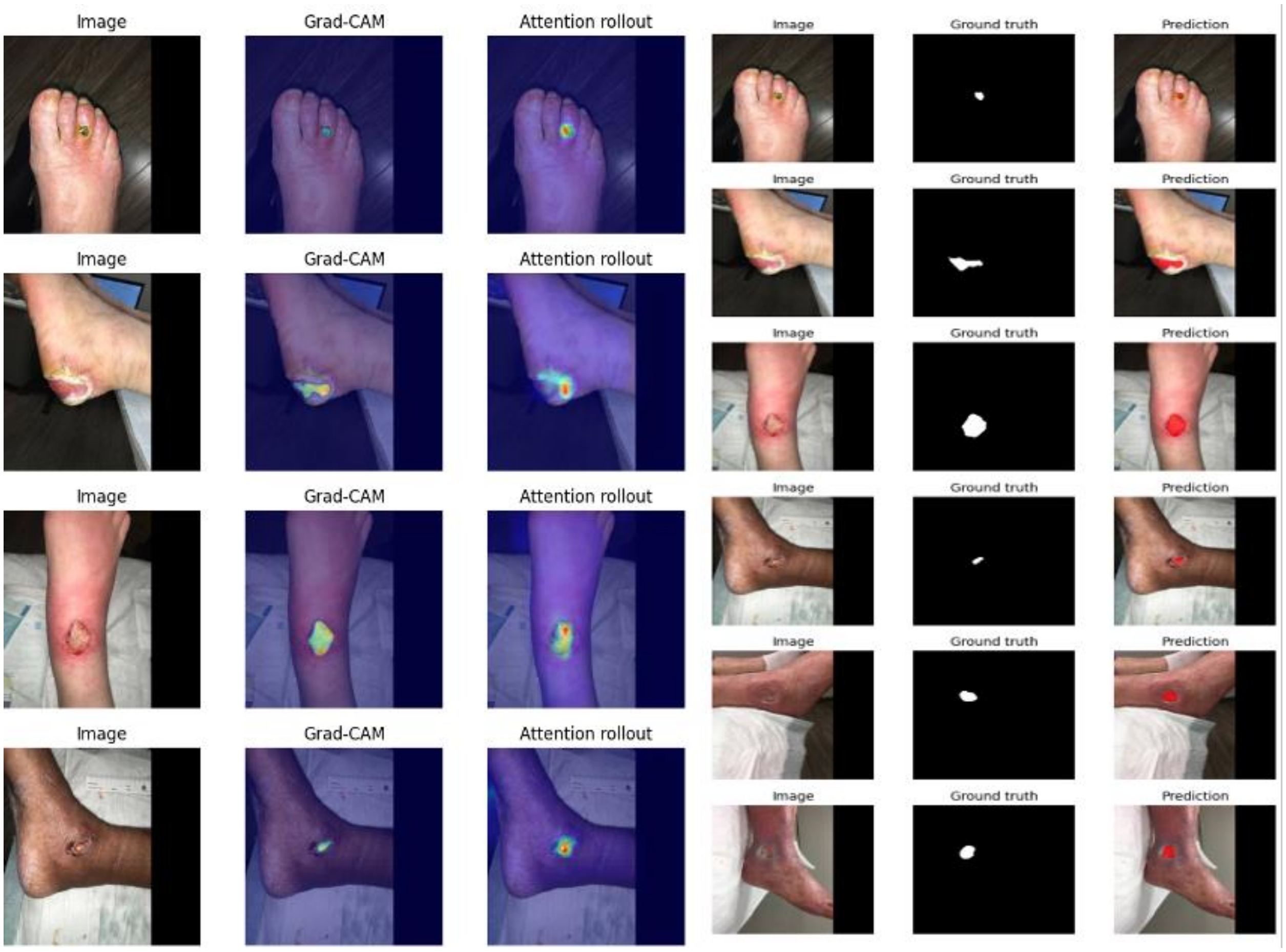

*Figure 7: Grad-CAM and attention-rollout explanations compared against ground-truth masks.*

### 3.6 Clinical Utility Analysis

To assess clinical relevance, we compared predicted and expert wound areas (Figure 8) and examined model calibration and predictive uncertainty (Figure 9). We observed a strong linear correlation (Pearson r = 0.944, R-squared = 0.892), indicating good reliability in automated wound sizing. A Bland-Altman analysis showed a small mean bias of approximately 22 pixels, but the 95% limits of agreement were wide (approximately −1412 to +1457 pixels) and driven primarily by a small number of very large wounds. Consequently, the model is well suited to tracking relative change in wounds of typical size, whereas absolute area estimates for the largest ulcers are subject to substantial uncertainty. Calibration and clinical-utility metrics are summarized in Table 9. The pixel-level expected calibration error is low (ECE = 0.0024), but this value should be interpreted with care: it is computed over all pixels, and under the strong foreground-background imbalance of wound photographs it is dominated by high-confidence background pixels that the model classifies almost perfectly. The reliability diagram in Figure 9 makes this explicit; predictions in the highest-confidence bin lie essentially on the diagonal, whereas the smaller population of intermediate-confidence pixels, which correspond predominantly to the wound boundary, is overconfident. The model is therefore well calibrated in aggregate but less well calibrated in the boundary region that most affects wound-area estimation. Consistent with this, predicted per-image uncertainty showed no significant relationship with per-image Dice (r = 0.029, p = 0.68), indicating that the model's uncertainty estimates do not reliably flag its own failures; we treat this as a limitation rather than as evidence of reliability. We note that calibration is a property of the predicted probabilities and is therefore reported separately from the operating-point selection described in Section 3.3.

*Table 9: Calibration and clinical-utility metrics for the full model.*

| Metric | Value |
|---|---|
| Expected Calibration Error (ECE), all pixels | 0.0024 (background-dominated; see Section 3.6) |
| Uncertainty vs Dice correlation | r = 0.029, p = 0.68 (no predictive relationship) |
| Wound-area agreement (Pearson r) | 0.944 |
| Wound-area agreement (R-squared) | 0.892 |
| Bland-Altman bias / 95% LoA | approx 22 px / -1412 to +1457 px |

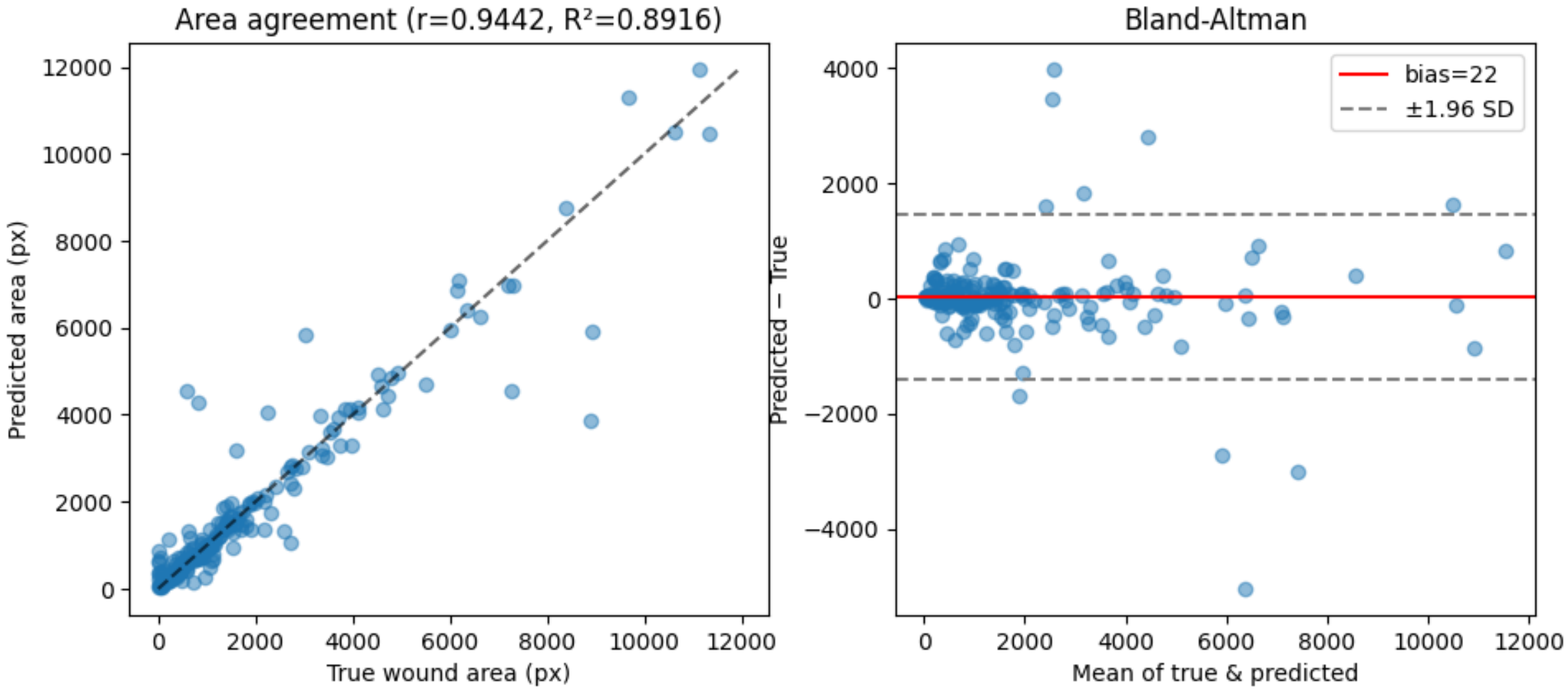

*Figure 8: Clinical-utility analysis. (A) Predicted versus expert wound area; (B) Bland-Altman agreement.*

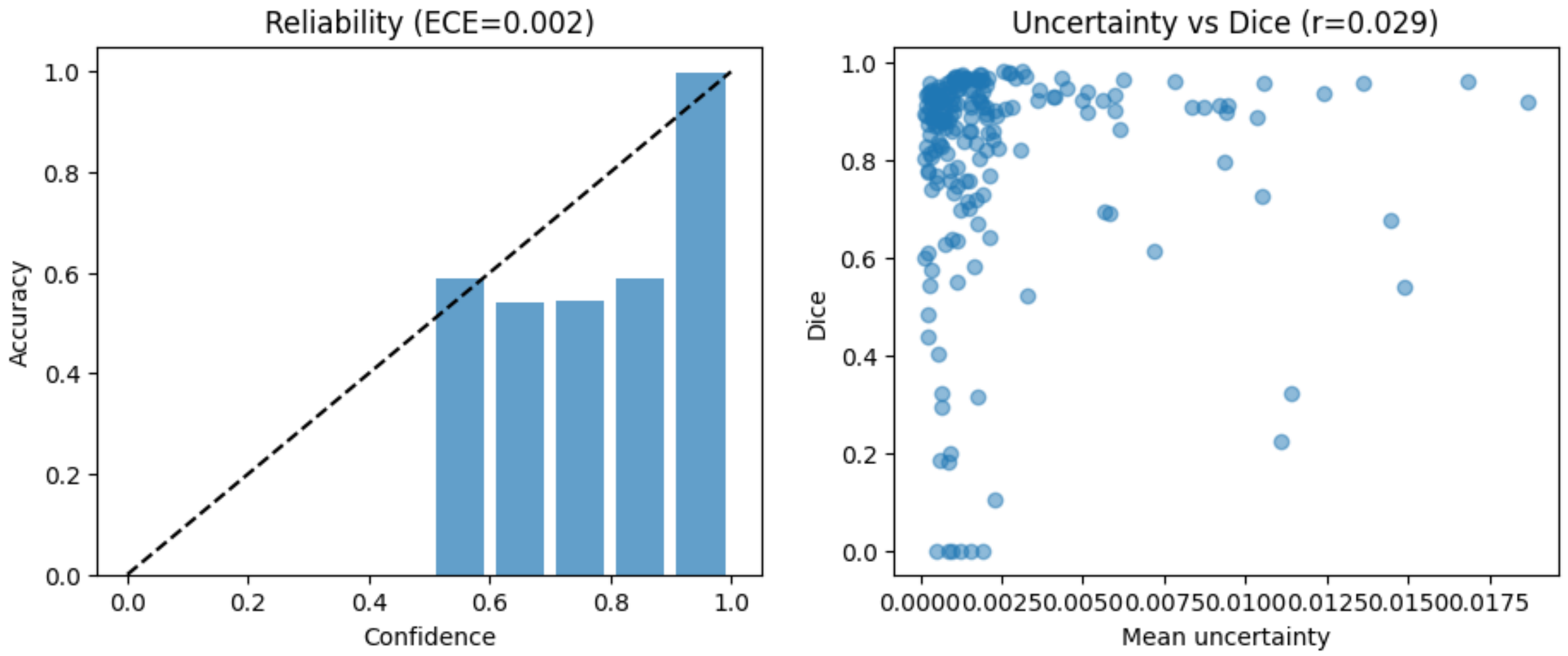

*Figure 9: Model calibration and predictive uncertainty. (A) Reliability diagram over all validation pixels (ECE = 0.0024); the dashed diagonal denotes perfect calibration. The highest-confidence bin contains the large majority of pixels and lies close to the diagonal, whereas intermediate-confidence bins fall below it, indicating overconfidence. (B) Mean per-image predictive entropy versus per-image Dice (r = 0.029, p = 0.68).*

### 3.7 Comparative Analysis

To contextualize the performance of the proposed framework against contemporary medical image segmentation techniques, we compiled reported results for recent architectures spanning 2024 to 2026 (Table 10), with publication years, evaluation datasets, explainability capabilities, and computational costs indicated for each entry. Recent prior work in this domain includes deep hybrid convolutional frameworks, dual-attention variants, lightweight state-space models, and integrated multi-scale transformer classification networks. Because these methods were evaluated on different corpora and under differing data splits, input resolutions and post-processing procedures, the compiled values are presented strictly for context and do not constitute a controlled ranking; the evaluation dataset for each entry is stated explicitly so that the reader can judge which comparisons are meaningful.

As shown in Table 10, the proposed model attains an internal validation Dice of 0.8035 on the FUSeg validation split. This is lower than the headline Dice reported for most of the methods listed, and we make no claim of competitive or state-of-the-art regional overlap. The comparison must, however, be read with care, because the entries were not evaluated on a common benchmark: the reported figures were obtained on chronic-wound photographs, on mixed acute and chronic wound cohorts, and on thermographic images, as well as under different data splits, input resolutions and post-processing procedures, so the values are not directly commensurable. Restricting attention to methods evaluated on FUSeg or on FUSeg-derived data narrows the spread considerably, with reported Dice values of 0.7363, 0.8340 and 0.9223 against the 0.8035 obtained here; on this benchmark the present model is therefore mid-range in overlap accuracy rather than leading. We report the wider comparison for transparency rather than to claim a favourable ranking, and the contribution of this work is accordingly not raw overlap accuracy. It lies instead in a gap that is common across these contemporary models: the absence of decision transparency and of any quantitative verification of attribution faithfulness. Several of the top-performing methods explicitly record the lack of explainability as an unresolved constraint or defer it to future work, and many report no parameter or computational-complexity budget, which is a decisive consideration for deployment in resource-constrained clinical settings.

By contrast, the proposed pipeline is a compact framework (8.79 M parameters and 28.77 GMACs) that natively embeds quantitative, multi-method explanation. Rather than relying solely on post-hoc convolutional gradients, our approach quantifies and reports both Grad-CAM and bottleneck attention-rollout pathways, and evaluates them against ground-truth lesions for localization and against the model's own output for faithfulness. Clinicians are thereby provided with dual, complementary attribution maps together with an explicit account of how far each reflects the underlying computation. The design philosophy of this work is accordingly to prioritise transparent reasoning, boundary-aware evaluation and multi-seed reproducibility rather than incremental gains in raw overlap.

*Table 10: Comparison with recent (2024 to 2026) diabetic foot ulcer and wound segmentation methods. Reported Dice values are taken from the original publications and were obtained on different datasets, splits, input resolutions and post-processing procedures, as indicated in the Dataset column; the table is therefore contextual and is not a like-for-like ranking.*

| **Method** | **Year** | **Approach** | **Dataset** | **Dice** | **Explainability (XAI)** | **Inference Time / Computational Complexity** |
|---|---|---|---|---|---|---|
| ResNet-U-Net Hybrid [24] | 2024 | ResNet34 encoder + U-Net decoder | Chronic Wound | 0.986 | None (noted as an explicit limitation) | Not reported |
| Dual Attention U-Net [22] | 2025 | VGG16 backbone + U-Net + dual attention | DFU + acute/chronic | 0.941 | Not included | Not reported |
| DFU-MambaLite UNet [12] | 2025 | Mobile U-Net framework + Mamba State Space Model | FUSC 2021 (FUSeg) | 0.922 | None (planned for future work) | 0.0878 seconds (2.34 GFLOPs) |
| SegFormer Ensemble [43] | 2025 | 3× MiT-B5 Ensemble + Weighted Soft Mean | FUSeg | 0.7363 | Not included | Heavy (3× 82M parameters; noted as unsuitable for edge devices) |
| U-MST + Capsule Network | 2026 | Modified Swin Transforme | Thermography (DFU) | 0.986 | None (planned for future work) | 0.37 seconds per image |
| [44] | | r + Capsule Network | | | | |
| SegFormer-B2 [45] | 2026 | Vision Transformer (MiT-B2) | FUSeg + AZH (in-domain) | 0.8340 | Not included | Not reported |
| TransUNet-GradCAM (**Ours**) | 2026 | Hybrid Transformer + quantitative XAI | FUSeg (validation) | 0.8035 | Integrated (Complementary Grad-CAM & Attention Rollout validation) | Lightweight (8.79 M parameters / 28.77 GMACs) |

### 3.8 External Validation

We evaluated generalization on two external cohorts from the Wound Seg 3 collection, applying the model directly with the fixed validation threshold and no fine-tuning (Table 11). On the AZH Wound Care Center cohort (n = 278) [1,6], the model achieved a Dice of 0.7460, IoU 0.6546, precision 0.859, recall 0.704, HD95 20.24 px and ASSD 6.72 px. Relative to the internal Dice of 0.8035, this is a drop of only 0.061, meaning the model retained approximately 92% of its in-domain Dice on an entirely separate clinical cohort [1], which we interpret as meaningful but partial generalization. On the small Medetec subset (n = 8) [26], the model reached a Dice of 0.5198 with very high precision (0.991) but low recall (0.377); with only eight images this is not a reliable benchmark, and we report it solely as a qualitative observation that the model under-segments on that domain. Qualitative external overlays are shown in Figure 10. Taken together, these results indicate partial transfer under domain shift; the characterization of the model as exhibiting robust zero-shot transferability, used in the previous version of this manuscript, is therefore not supported and has been withdrawn.

*Table 11: External validation under domain shift (single fixed threshold, no fine-tuning).*

| Dataset | n | Dice | IoU | Precision | Recall | HD95 | ASSD |
|---|---|---|---|---|---|---|---|
| AZH (Wound Seg 3) | 278 | 0.7460 | 0.6546 | 0.859 | 0.704 | 20.24 | 6.72 |
| Medetec (Wound Seg 3) | 8 | 0.5198 | 0.3741 | 0.991 | 0.377 | 59.4 | 22.4 |

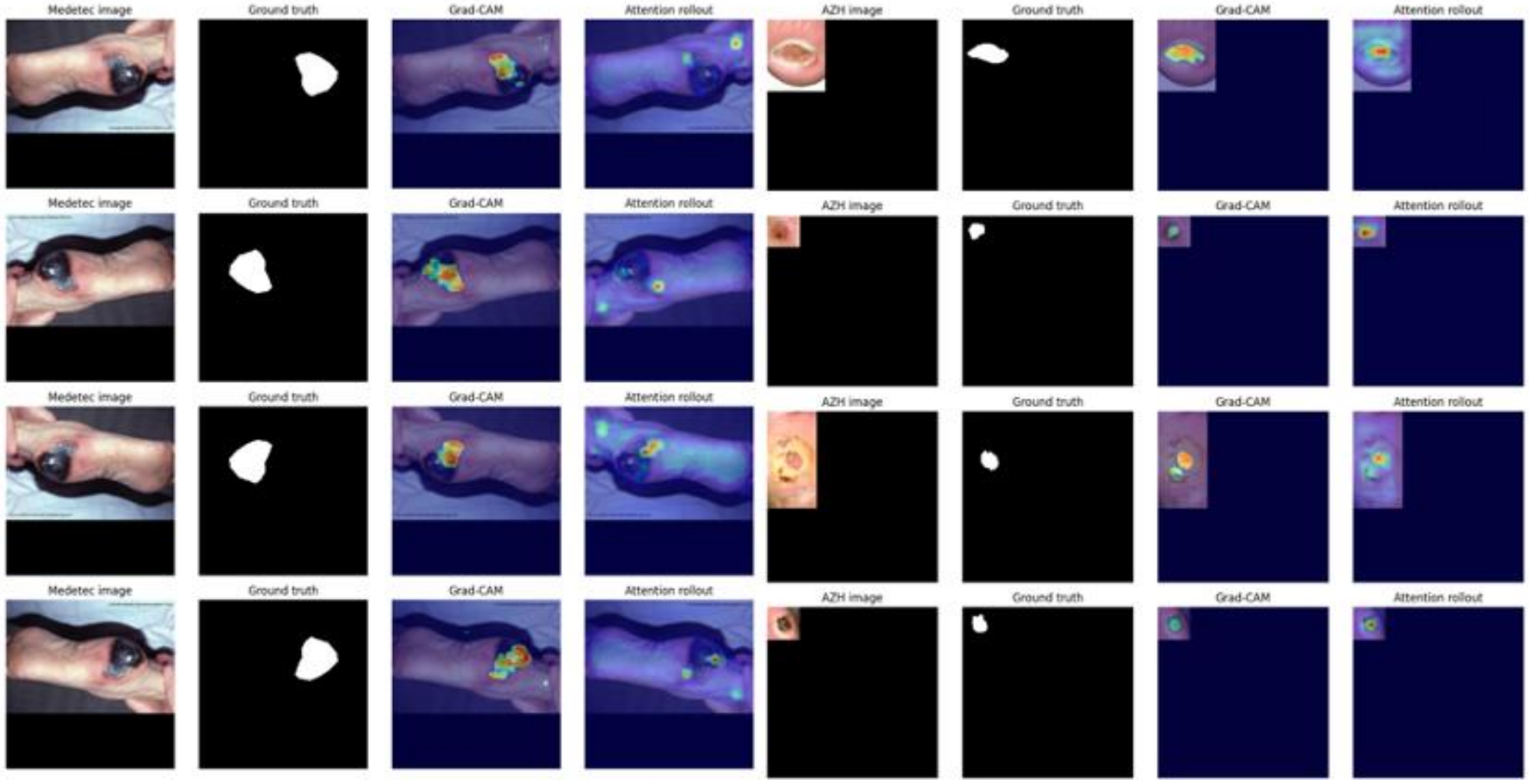

*Figure 10: Qualitative external-validation results on AZH and Medetec, with Grad-CAM and attention-rollout overlays.*

## 4. Discussion

The results indicate that the hybrid ViT-bottleneck U-Net is a reliable framework for diabetic foot ulcer segmentation, though it does not exceed the strongest reported methods. The internal Dice of 0.8035 ± 0.0053, reported over five seeds with boundary-aware metrics, is competitive with recent baselines [12,13,45], and its value lies less in its absolute magnitude than in being reported with full statistical context and a clear account of what each component contributes. The small between-seed variance shows that the result is reproducible rather than the product of a single fortunate run.

Our ablation clarifies the source of performance. Of the components examined, only the hybrid loss produced a statistically significant change in Dice, which is consistent with the known benefit of combining a region-overlap term with cross-entropy under severe class imbalance [38]. The Transformer bottleneck, by contrast, did not significantly improve internal Dice; we therefore do not attribute in-domain accuracy to it. Its justification in this work is interpretability, since it enables the attention-rollout analysis that we found to be the more faithful explainer [42] , together with its established role in modelling global context. Being explicit about this distinction is, in our view, more scientifically useful than claiming uniform benefit from every component.

The explainability analysis also speaks to a common concern with Grad-CAM on Transformer models [25]. Because Grad-CAM reads convolutional activations, it can appear well-localized yet fail to reflect the attention mechanism driving the model. By quantifying faithfulness, we showed that attention rollout, while less localized, is significantly more faithful, and we therefore report both rather than relying on Grad-CAM alone.

Clinically, the strong area agreement (r = 0.944) supports the use of the model for longitudinal wound tracking [46], while the wide Bland-Altman limits for large wounds indicate that absolute area estimates in those cases should be treated with caution. The external results reinforce a measured interpretation: strong retention on the larger Zenith of Healthcare (AZH) Wound Care Center cohort and under-segmentation on the tiny Medetec subset together indicate partial generalization under domain shift rather than robust zero-shot transfer [20].

### 4.1 Limitations and Future Work

Several limitations remain. First, external performance, while strong on AZH, declined on the small Medetec subset, so generalization is only partial, and the model is not yet validated for deployment across arbitrary clinical settings. Second, although we segment at 384 by 384, higher-resolution inputs might capture fine texture detail relevant for tissue sub-typing; we did not explore this owing to memory constraints. Third, our ablation shows that several architectural components do not significantly improve in-domain Dice, so part of the model's capacity may be unnecessary for accuracy alone and is retained mainly for interpretability and external robustness. Fourth, the Bland-Altman limits of agreement are wide for very large wounds, so absolute area estimates in those cases require caution. Fifth, the external evidence is itself limited in scale: the AZH cohort contains 278 images and the Medetec subset only 8, so the latter supports qualitative observation rather than a reliable quantitative estimate of performance. Sixth, because the official FUSeg test split is unlabeled, all internal quantitative evaluation rests on the 200-image validation set, which constrains the precision of the internal performance estimates. Seventh, the model was trained on a single source dataset (FUSeg); this single-source training is a likely contributor to the sensitivity observed under domain shift, and multi-source training may improve robustness. Eighth, calibration was assessed at the pixel level and is dominated by the large background class, so the low aggregate ECE overstates calibration quality at the wound boundary, where intermediate-confidence predictions are overconfident; in addition, predicted uncertainty did not correlate with per-image Dice, so the model's confidence should not be used as a failure detector in its present form. Future work will expand external validation to larger and more diverse cohorts, investigate higher-resolution training, and explore whether the Transformer bottleneck can be simplified without loss of interpretability.

## 5. Conclusion

We presented a hybrid ViT-bottleneck U-Net for diabetic foot ulcer segmentation and, more importantly, a rigorous and transparent evaluation of it. On the FUSeg validation set the model achieved a Dice of $0.8035 \pm 0.0053$ and an IoU of $0.7149 \pm 0.0073$ over five seeds, a result that is competitive with recent baselines rather than a new state of the art. A component-wise ablation identified the hybrid loss as the only component with a statistically significant effect on in-domain Dice, and a quantitative explainability analysis showed that Grad-CAM and attention rollout are complementary, with rollout being significantly more faithful. External validation on the AZH cohort retained about 92% of internal Dice, while the small Medetec subset revealed a domain in which the model under-segments; we therefore conclude that the framework generalizes partially rather than robustly under domain shift.

The primary contributions of this work are methodological: multi-seed statistics with confidence intervals and significance tests, a component-wise ablation, boundary-aware metrics, a faithfulness-based explainability comparison, and an honest external-validation study. We believe this transparent style of reporting is of greater value to the automated wound-assessment community than an incremental gain in headline accuracy. At the same time, the limitations set out above, in particular the partial external generalization, the small external test sets, and the single-source training, temper any clinical claims: the framework should be regarded as a reproducible and interpretable research baseline rather than a deployment-ready tool, and broader multi-centre validation would be required before clinical use.

## List of Abbreviations

ASSD: Average Symmetric Surface Distance
AUC: Area Under the Curve
AZH: Advancing the Zenith of Healthcare (Wound Care Center dataset)
BCE: Binary Cross-Entropy
BN: Batch Normalization
CI: Confidence Interval
CNN: Convolutional Neural Network
DFU: Diabetic Foot Ulcer
ECE: Expected Calibration Error
FN: False Negative
FP: False Positive
FUSeg: Foot Ulcer Segmentation (Challenge)
GMACs: Giga Multiply-Accumulate Operations
Grad-CAM: Gradient-weighted Class Activation Mapping
HD95: 95th-percentile Hausdorff Distance
HIPAA: Health Insurance Portability and Accountability Act
IoU: Intersection over Union
LoA: Limits of Agreement
MHA: Multi-Head Attention
MLP: Multi-Layer Perceptron
PR: Precision-Recall
ReLU: Rectified Linear Unit
ROC: Receiver Operating Characteristic
TN: True Negative
TP: True Positive
ViT: Vision Transformer

## Declarations

### Ethics approval and consent to participate

This study used publicly available, fully de-identified image datasets (the Foot Ulcer Segmentation Challenge dataset and the publicly released AZH and Medetec wound datasets). All images were de-identified at source in accordance with HIPAA requirements, and no personally identifiable information was accessed. Because the work relies solely on these publicly available, de-identified datasets and did not involve any new collection of data from human participants, no additional institutional ethical approval was required. The study was conducted in accordance with the principles of the Declaration of Helsinki and all relevant data-privacy standards.

**Consent for publication**
Not applicable. This manuscript does not contain any individual person's identifiable data in any form.

**Availability of data and materials**
The datasets analysed during the current study are publicly available. The Foot Ulcer Segmentation (FUSeg) Challenge dataset is a public, open-access resource accessible upon registration at https://fusc.grand-challenge.org, and the AZH and Medetec wound datasets are available via https://www.kaggle.com/datasets/adnanjan01/wound-seg-3-datasets

**Competing interests**
The authors declare that they have no competing interests.

**Funding**
This research received no external funding. The work was conducted using institutional resources and facilities.

**Authors' contributions**
A.A. conceived the study, designed and implemented the model, conducted the experiments and analysis, and drafted the manuscript. M.S. and J.W.A. reviewed and edited the manuscript. S.E.M. supervised the work and reviewed and edited the manuscript. All authors read and approved the final manuscript.

**Acknowledgements**
The authors thank the organizers of the Foot Ulcer Segmentation Challenge and the providers of the AZH and Medetec wound datasets for making their data publicly available.